\documentclass[10pt,a4paper]{article}
\usepackage{epsfig}
\usepackage{amsmath}
\usepackage{amssymb}

\newcommand{\institute}[1]{\parbox{16cm}{%
\centering\normalsize \sl #1}}
\newcommand{\Prsu}{\mathop\mathrm{Proj}\nolimits_{SU(3)}}
\newcommand{\Tr}{\mathop \mathrm{Tr}}
\newcommand{\Ree}{\mathop \mathrm{Re}}
\newcommand{\D}{\displaystyle}
\newcommand{\Lwab}{{\mathcal{L}_{w^2}^2
\left(\left[a,b\right]\right)}}
\newcommand{\Lab}{{\mathcal{L}^2\left(\left[a,b\right]\right)}}
\newcommand{\scalar}[2]{{\ensuremath \left< #1,#2\right>}}
\newcommand{\scalarw}[2]{{\ensuremath \left< #1,#2\right>_w}}
\newcommand{\ketbra}[2]{{\ensuremath \left| #1 \right> \!\left< #2
\right|}} 
\newcommand{\ketbraw}[2]{{\ensuremath \left| #1 \right> \!\left< #2
\right|_w}}

\newcommand{\normw}[1]{\ensuremath \left\| #1 \right\|_w}
\newcommand{\Int}[4]{\displaystyle \mathop \int_{#3}^{#4} #1
\,\mathrm{d}#2} 
\newcommand{\id}{\mathop{\mathrm{id}}\nolimits}
\newcommand{\ssum}{\mathop \mathrm{(s)\!\!}\sum\limits}
\newcommand{\Sim}[1]{\int_{#1,\text{num.}}}
\newcommand{\Frac}[2]{\frac{\displaystyle #1}{\displaystyle #2}}

\begin{document}

\title{{\normalsize \vspace*{-2cm}\hfill\mbox{WUB 03-13}\\
\hfill\mbox{ITP-Budapest 604}\\ \vspace*{2cm}}
Least-Squared Optimized Polynomials for Smeared Link Actions}
\date{today}

\author{
S.~D.~Katz$^a$\thanks{On leave from Institute for
Theoretical Physics, E\"otv\"os University,
Budapest, Hungary.} \,and B.~C.~T\'oth$^{b}$\\
\institute{
$^a$Department of Physics, University of Wuppertal, Germany\\
$^b$Institute for Theoretical Physics, E\"otv\"os University, 
P\'azm\'any 1/A, H-1117 Budapest, Hungary}}
\date{\today}

\maketitle

\begin{abstract}
  We introduce a numerical method for generating the
  approximating polynomials used in fermionic calculations with
  smeared link actions. We investigate the stability of the algorithm
  and determine the optimal weight function and the optimal type of
  discretization. The achievable order of polynomial approximation
  reaches several thousands allowing fermionic calculations using
  the Hypercubic Smeared Link action even with
  physical quark masses.
\end{abstract}

\section{Introduction}
\label{sec:intro}
The usage of smeared or fat links improves flavor symmetry for
staggered fermions~\cite{9809157,9903032,0103029}.  
Since the smearing contains a
projection onto $SU(3)$, the Hypercubic Smeared Link (HYP) action is
not bilinear in the original thin link variables. Therefore, the
explicit form of the fermion force is rather complicated making the
Hybrid Monte-Carlo (HMC) and other molecular dynamics based algorithms
with the HYP action practically unusable\footnote{Note, that using a
  different projection one can calculate the fermion
  force~\cite{Kamleh:2003wb}. This idea was applied for the fat link
  irrelevant clover (FLIC) action~\cite{Zanotti:2001yb}. Recently
  a completely analytic smearing without any projection has also 
  been introduced~\cite{Morningstar:2003gk}.}. An update
method based on a stochastic estimator \cite{0012022,0203010} can
avoid this problem.  The algorithm using improved stochastic
estimators \cite{0203026} requires polynomial approximation of
functions of type $x^{-\alpha}e^{p(x)}$, where $\alpha > 0$ and $p(x)$
is a low order polynomial. When the calculations are made at the small
physical quark masses the order of these polynomials have to be in the
range of the thousands. We introduce a numerical method to generate
these high order polynomials and investigate the stability, optimal
weight function and optimal type of discretization for the algorithm.
In contrast to exact methods our procedure is very fast, stable up to
thousands of orders and can be applied practically to functions of any
type.

The rest of the paper is organized as follows. In the next
section we summerize the properties of the HYP action. In
Sect.~\ref{sec:poly} we describe the process of generating the
approximating polynomials. We conclude in Sect.~\ref{sec:conc}.

\section{The HYP action}
\label{sec:HYP}

The construction of the Hypercubic Smeared Link (HYP) action goes as
follows \cite{0012022,0203026}.  First, the original thin links
$U_{i,\mu}$ are used to construct the set of decorated fat links,
$V_{i,\mu;\nu\rho}$ with a modified projected APE blocking step
\begin{equation} 
V_{i,\mu;\nu\rho} := \Prsu
\left[(1-\alpha_3)U_{i,\mu} + \frac{\alpha_3}{2}
\sum_{\pm\eta\ne\rho,\nu,\mu} U_{i,\eta} U_{i+\hat{\eta},\mu}
U_{i+\hat\mu,\eta}^{\dagger}\right].
\label{eqn:step1}
\end{equation}
The indices $\nu$ and $\rho$ indicate that the fat link
$V_{i,\mu;\nu\rho}$ in direction $\mu$ is not decorated with staples
extending in directions $\nu$ and $\rho$. The projection to $SU(3)$
can be defined in two different ways. $W\in SU(3)$ is the
\emph{deterministic} projection of $A$,
\begin{equation}
W=\Prsu A \quad\text{if} \quad \Ree\Tr\left(WA^{\dagger}\right) =
\max_{U\in SU(3)} \Ree\Tr \left(UA^{\dagger}\right),
\label{eqn:dprojsu3}
\end{equation}
whereas the \emph{probabilistic} projection $W^\lambda$ of $A$ is
chosen according to the probability distribution
\begin{equation}
P(W)\propto
\exp\left[\frac{\lambda}{3}\Ree\Tr\left(WA^\dagger\right)\right].
\label{eqn:pprojsu3}
\end{equation}
In the second step the decorated links $\tilde{V}_{i,\mu;\nu}$ are
constructed using the fat links $V_{i,\mu;\nu\rho}$ obtained in the
first step as
\begin{equation}
\tilde{V}_{i,\mu;\nu} := \Prsu\left[ (1-\alpha_2)U_{i,\mu} +
\frac{\alpha_2}{4} \sum_{\pm\rho\ne\nu,\mu} V_{i,\rho;\nu\mu}
V_{i+\hat{\rho},\mu;\rho\nu} V_{i+\hat{\mu},\rho;\nu\mu}^\dagger
\right].
\label{eqn:step2}
\end{equation}
In the final step the blocked links $V_{i,\mu}$ are constructed as
\begin{equation}
V_{i,\mu}:=\Prsu\left[ (1-\alpha_1)U_{i,\mu} + \frac{\alpha_1}{6}
\sum_{\pm\nu\ne\mu} \tilde{V}_{i,\nu;\mu}
\tilde{V}_{i+\hat{\nu},\mu;\nu} \tilde{V}_{i+\hat{\mu},\nu;\mu}^\dagger
\right].
\label{eqn:step3}
\end{equation}
The smeared link obtained using the above construction containes thin
links only from hypercubes attached to the original thin link.

The HYP action is of the form
\begin{equation}
S=S_g(U)+\bar{S}_g(V)+S_f(V),
\label{eqn:hypaction}
\end{equation}
where $S_g(U)$ is the plaquette gauge action
\begin{equation}
S_g(U)=-\frac{\beta}{3}\sum_p\Ree\Tr(U_p)
\end{equation}
depending on the thin links $\{U\}$ and $\bar{S}_g(V)$ is the gauge
action depending on the smeared links $\{V\}$ \cite{0203026}. The main
role of $\bar{S}_g(V)$ is to increase the acceptance rate in the
accept-reject step. The simplest choice is the smeared plaquette
$\bar{S}_g(V)=-\frac{\gamma}{3}\sum_p\Ree\Tr(V_p)$, where $\gamma$ can
be tuned to maximize the acceptance rate. The fermions are coupled to
the smeared links, thus, the staggered fermionic matrix is of the form
\begin{equation}
M(V)_{i,j}=2m\delta_{i,j}+ \sum_\mu \eta_{i,\mu}\left( V_{i,\mu}
\delta_{i,j-\hat{\mu}} - V_{i-\hat{\mu},\mu}^\dagger
\delta_{i,j+\hat{\mu}} \right).
\label{eqn:staggeredmatrix}
\end{equation}
The matrix $M(V)^\dagger M(V)$ is block diagonal on even and odd
lattice sites. Let $\Omega$ denote the even block
\begin{equation}
\Omega(V) := \left( M(V)^\dagger M(V)
\right)_{\text{even},\text{even}}.
\end{equation}
Then the fermionic action $S_f(V)$ describing $n_f$ flavors is of the
form
\begin{equation}
S_f(V)=-\frac{n_f}{4} \Tr\ln \Omega(V).
\end{equation}

Since the dependence of the smeared links $\{V\}$ on the thin links
$\{U\}$ is nonlinear due to the projections to $SU(3)$, the explicit
form of the fermionic force, which is needed for molecular dynamics
simulations, is very complicated. This fact makes the HMC and R
algorithms virtually unusable. A two step algorithm, the partial
global stochastic Metropolis update is used instead.
In the first step a subset of the thin links $\{U\}$ is updated such
that the detailed balance condition with the thin link gauge action
$S_g(U)$ is satisfied. This can be done using either heatbath or
overrelaxation. In the second step the new smeared links $\{V'\}$ are
computed and the newly obtained configuration is accepted with the
probability
\begin{equation}
P_{\text{acc}}=\min \left\{1, \exp\left[
-\bar{S}_g(V')+\bar{S}_g(V)\right] \frac{\det \Omega(V')}{\det
\Omega(V)} \right\}.
\end{equation}
Instead of calculating the ratio of the determinants a stochastic
estimator is used. The ratio can be expressed as an expectation value
\begin{eqnarray}
\frac{\det \Omega(V')}{\det\Omega(V)} &=& \frac{\D
\int d\xi\xi^* \exp\left(-\xi^*\Omega(V')^{-1}\Omega(V)\xi\right)}
{\D \int d\xi\xi^* \exp(-\xi^*\xi)} \nonumber \\
&=& \left< \exp\left(-\xi^*\left[\Omega(V')^{-1}\Omega(V)-1\right]
\xi\right)\right>_{\xi^*\xi}.
\end{eqnarray}
Only one random source $\xi$ is used on every gauge configuration pair
$\{U\}$ and $\{U'\}$ to estimate the determinant ratio. The
expectation value is taken together with the configuration ensemble
average. Then the stochastic acceptance probability becomes
\begin{equation}
P_{\text{stoch}}=\min\left\{ 1,
\exp\left(-\bar{S}_g(V')+\bar{S}_g(V)\right)
\exp\left(-\xi^*\left[\Omega(V')^{-1}\Omega(V)-1\right]
\xi\right)\right\}.
\end{equation}

If the stochastic estimator has large fluctuations then the acceptance
rate can be very small even if the old and new fermionic determinants
are almost the same. The standard deviation of the stochastic
estimation
\begin{equation}
\left< \exp\left(-\Delta S_f\right) \right> = \frac{\det \Omega'}{\det
\Omega} = \det(A)^{-1} = \left<\exp \left(-\xi^* \left[A-1 \right] \xi
\right) \right>_{\xi^*\xi}
\end{equation}
can be written in the form
\begin{eqnarray}
\sigma^2 &=& \left< \exp\left(-2\xi^*\left[A-1\right] \xi\right)
\right>_{\xi^*\xi} - \left< \exp\left(-\xi^*\left[A-1\right] \xi\right)
\right>_{\xi^*\xi}^2 \nonumber \\
&=& \det\left(2A-1\right)^{-1}-\det(A)^{-2},
\label{eqn:std}
\end{eqnarray}
where $\Omega=\Omega(V)$, $\Omega'=\Omega(V')$ and
$A=\Omega'^{-1}\Omega$.
As suggested in \cite{Hasenbusch:1998yb,deForcrand:1998sv}, instead of $\Omega$ and $\Omega'$ we introduce 
the reduced matrices 
\begin{equation}
\Omega_r=\Omega e^{-2f(\Omega)},\quad\quad \Omega'_r=\Omega'
e^{-2f(\Omega')},
\end{equation}
where $f$ is a polynomial chosen such that the
function $e^{2f(x)}/x$ is close to 1 in the interval where the
eigenvalues of the matrix $\Omega$ can be found.
Then the ratio of the determinants can be rewritten as
\begin{eqnarray}
\frac{\det \Omega'}{\det \Omega} &=& \frac{\det \Omega'_r}{\det
\Omega_r} \exp\left( 2 \Tr \left[f(\Omega')-f(\Omega)\right]\right)
\nonumber \\
&=&\left< \exp \left( -\xi^*\left[\Omega_r'^{-1}\Omega_r - 1\right]
\xi \right) \right>_{\xi^*\xi} \exp\left( 2 \Tr
\left[f(\Omega')-f(\Omega)\right]\right).
\end{eqnarray}
Since the second factor can be evaluated exactly only the first factor
has to be evaluated stochastically.  Due to the special choice of $f$,
$A_r=\Omega_r'^{-1}\Omega_r \approx 1$, so the fluctuations of the
stochastic estimator are minimized, improving the acceptance rate.

Equation (\ref{eqn:std}) is valid for the standard deviation of the
stochastic estimator only if the matrix $2A-1$ is positive definite,
that is, all eigenvalues of $A$ are greater than $1/2$. This is,
however, very unlikely if the updating method changes a large number
of links of the configuration. In order to avoid this problem the
reduced fermionic determinant ratio can be written in the form
\begin{equation}
\label{eqn:detar}
\det(A_r)^{-1} = \det\left(A_r^{1/n}\right)^{-n} = \left<\exp \left(
-\sum_{j=1}^{n} \xi_j^* \left[ A_r^{1/n}-1 \right] \xi_j\right)
\right>_{\xi_j^*\xi_j},
\end{equation}
where $n$ is an arbitrary positive integer and $\xi_j$ are $n$
independent random vectors. Then the standard deviation becomes
\begin{equation}
\sigma^2=\det\left(2A_r^{1/n}-1\right)^{-n}-
\det\left(A_r\right)^{-2}.
\end{equation}
This is valid only if all the eigenvalues of $A_r$ are greater than
$2^{-n}$. If $n$ is chosen large enough this condition can always be
fulfilled. Since the determinant of a matrix product is independent of
the order of the matrices, the $n$th root of $A_r$ can be written as
\begin{equation}
A_r^{1/n} = \Omega_r'^{-1/2n}\Omega_r^{1/n}\Omega_r'^{-1/2n}.
\end{equation}
The factors can be approximated by polynomials as
\begin{eqnarray}
\Omega_r'^{-1/2n} &=& \Omega'^{-1/2n}\exp\left(f(\Omega')/n\right) =
P_l^{(2n)}(\Omega'), \nonumber \\
\Omega_r^{1/n} &=& \Omega^{1/n} \exp \left( -2f(\Omega)/n \right) =
Q_k^{(n)}(\Omega).
\end{eqnarray}
Here $P_l^{(2n)}$ and $Q_k^{(n)}$ are $l$ and $k$ order polynomials of
the fermionic matrices $\Omega$ and $\Omega'$, respectively. Then all
the terms of the exponent of (\ref{eqn:detar}) can be written in the
form
\begin{equation}
\xi^*\left[A_r^{1/n}-1\right]\xi = \xi^* P_l^{(2n)}(\Omega')
Q_k^{(n)}(\Omega) P_l^{(2n)}(\Omega') \xi - \xi^* P_l^{(2n)}(\Omega')
Q_k^{(n)}(\Omega') P_l^{(2n)}(\Omega')\xi.
\end{equation}
The polynomial orders $l$ and $k$ required for a reasonable
approximation depend on the used quark mass. The polynomials should be
optimized for the interval spanned by the eigenvalues of $\Omega$. 
The smallest possible eigenvalue is $4m^2$, so if smaller and smaller
quark masses are used then higher order polynomials are
required. The polynomials have to be generated only once before each
simulation using the method described in the following section.

\section{Generating the polynomials}
\label{sec:poly}

Our aim is to approximate the function $f$ in the interval
$\left[a,b\right]$ using an $n$th order polynomial $P_f^{(n)}(x)$. We
choose a weight function $w(x)$ and define the deviation of
$P_f^{(n)}(x)$ from $f$ using the distance in the Hilbert space $\Lwab$
as
\begin{equation}
\delta_n = \frac{\normw{f-P^{(n)}_f}}{\normw{f}} =
\Frac{\sqrt{\scalarw{f-P_f^{(n)}}{f-P_f^{(n)}}}}{\sqrt{\scalarw{f}{f}}}= 
\Frac{\sqrt{\Int{\left| f(x)-P_f^{(n)}(x)\right|^2 w(x)^2}{x}{a}{b}}}
{\sqrt{\Int{\left| f(x)\right|^2 w(x)^2}{x}{a}{b}}}. 
\end{equation}
Here $\scalarw{}{}$ and $\normw{\,}$ denote the inner product
\begin{equation}
\scalarw{f}{g}=\Int{f(x)^* g(x) w(x)^2}{x}{a}{b}
\end{equation}
and the norm
\begin{equation}
\normw{f}=\sqrt{\scalarw{f}{f}}
\end{equation}
in the Hilbert space $\Lwab$, respectively. For the best choice of
$w(x)$ see Section \ref{sec:stab}. In order to minimize
$\delta_n$ we take a basis of orthogonal polynomials $\Phi_\mu$,
\begin{equation}
\label{eqn:q2}
\scalarw{\Phi_\mu}{\Phi_\nu}=\delta_{\mu\nu}q_\nu,
\end{equation}
where $\Phi_\mu$ ($\mu=0,1,2,\dots$) is a $\mu$th order polynomial
with norm
\begin{equation}
\label{eqn:q1}
q_\mu=\normw{\Phi_\mu}^2=\Int{\left|\Phi_\mu(x)\right|^2
w(x)^2}{x}{a}{b}.
\end{equation}
This basis of polynomials is generated using the Gram-Schmidt
orthogonalization process from the simple polynomials $\id^0\,(\equiv\!
1)$, $\id^1\,(\equiv\! x)$, $\id^2\,(\equiv\! x^2)$, \dots{}:
\begin{eqnarray}
\Phi_0 &:=& \id^0=1 \nonumber\\
\Phi_1 &:=&  \left( I - \frac{\ketbraw{\Phi_0}{\Phi_0}}
{\normw{\Phi_0}^2} \right)  \id^1=
\id^1-\frac{\scalarw{\Phi_0}{\id^1}}{q_0} \,\Phi_0 \nonumber\\
\Phi_2 &:=& \left( I - \frac{\ketbraw{\Phi_0}{\Phi_0}}
{\normw{\Phi_0}^2} - \frac{\ketbraw{\Phi_1}{\Phi_1}}
{\normw{\Phi_1}^2} \right)  \id^2=
\id^2-\frac{\scalarw{\Phi_0} {\id^2}} {q_0} \,\Phi_0
-\frac{\scalarw{\Phi_1}{\id^2}} {q_1} \,\Phi_1 \nonumber\\
\Phi_{\mu+1} &:=& \left( I - \sum_{\nu=0}^{\mu} \frac{\ketbraw{\Phi_\nu}
{\Phi_\nu}}{\normw{\Phi_\nu}^2} \right)  \id^{\mu+1}= \id^{\mu+1} -
\sum_{\nu=0}^{\mu} \frac{1}{q_\nu} \scalarw{\Phi_\nu}{\id^{\mu+1}}
\Phi_\nu.
\label{eqn:gram-schmidt}
\end{eqnarray}
Using this basis $f$ can be written as
\begin{equation}
\label{eqn:f=}
f=\sum_{\mu=0}^{\infty}
\frac{\scalarw{\Phi_\mu}{f}}{\scalarw{\Phi_\mu}{\Phi_\mu}} =
\sum_{\mu=0}^{\infty} \frac{b_\mu}{q_\mu} \Phi_\mu,
\end{equation}
where
\begin{equation}
\label{eqn:b}
b_\mu:= \scalarw{\Phi_\mu}{f}=\Int{\Phi_\mu(x)f(x)w(x)^2}{x}{a}{b}.
\end{equation}
The $n$th order polynomial $P_f^{(n)}$ for which $\delta_n$ is minimal
can be obtained by taking only the first $n$ terms of the sum in
(\ref{eqn:f=}).
\begin{equation}
\label{eqn:Pn}
P_f^{(n)}:= \sum_{\mu=0}^{n} \frac{b_\mu}{q_\mu} \Phi_\mu =
\sum_{\mu=0}^{n} c_\mu \Phi_\mu, \quad\quad c_\mu:= 
\frac{b_\mu}{q_\mu}
\end{equation}

In order to obtain the polynomials $\Phi_\mu$ a second order recursion
formula \cite{9903029,9911014} can be used instead of the numerically
unstable Gram-Schmidt orthogonalization process
(\ref{eqn:gram-schmidt}). The recursion goes as follows. The first two
polynomials are obtained identically according to (\ref{eqn:gram-schmidt})
\begin{equation}
\label{eqn:phi01}
\Phi_0(x)=1,\quad\quad \Phi_1(x)=x-\frac{\D \Int{w(x)^2x}{x}{a}{b}}
{\D \Int{w(x)^2}{x}{a}{b}}.
\end{equation}
Let
\begin{equation}
\label{eqn:p}
p_\mu:= \Int{\Phi_\mu(x)^2 w(x)^2x}{x}{a}{b}
\end{equation}
and
\begin{equation}
\label{eqn:betagamma}
\beta_\mu:=-\frac{p_\mu}{q_\mu}, \quad
\gamma_{\mu-1}:=-\frac{q_\mu}{q_{\mu-1}}.
\end{equation}
Then the rest of the polynomials can be obtained as
\begin{equation}
\label{eqn:rek}
\Phi_{\mu+1}(x)=(x+\beta_\mu)\Phi_\mu(x)+\gamma_{\mu-1}\Phi_{\mu-1} (x).
\end{equation}

\subsection{Numerical realization}
\label{sec:num}

One proper method of proceeding with the algorithm is to calculate the
integrals (\ref{eqn:q1}), (\ref{eqn:b}) and (\ref{eqn:p}) exactly.
This method is followed in Ref.~\cite{0302025}. These calculations
require a precision of several hundreds or thousands of digits which
can be carried out using multiprecision arithmetics libraries. Since
the integrals are carried out exactly the indefinite integrals of the
integrands have to be known. As a consequence only special types of
functions $f$ can be approximated and the weight function $w$ also has
to be carefully chosen.

The method we use is to calculate the integrals numerically. The
interval $\left[a,b\right]$ is divided into N subintervals (N+1
discretization points) logarithmically (see Section \ref{sec:stab}).
Since all the integrals are calculated using Simpson's rule, N has to
be even. The values of the function $f$, polynomials $\Phi_\mu$ and
the weight function $w^2$ are calculated and stored only at the
discretization points. First $\Phi_0$ and $\Phi_1$ are determined with
the integrals in (\ref{eqn:phi01}) carried out numerically. In the
$\mu$th step $q_\mu$ and $p_\mu$ are calculated first using
(\ref{eqn:q1}) and (\ref{eqn:p}), then $\beta_\mu$ and
$\gamma_{\mu-1}$ using (\ref{eqn:betagamma}). Then $\Phi_{\mu+1}$ is
determined at each discretization point from $\Phi_\mu$ and
$\Phi_{\mu-1}$ using (\ref{eqn:rek}). Then $b_\mu$ and $c_\mu$ are
calculated using (\ref{eqn:b}). Finally, $c_\mu\cdot \Phi_\mu$ is
added to the actual value of $P_f^{(n)}$.

This method has three major advantages. Firstly, we have a second
order recursion formula (\ref{eqn:rek}). Therefore, at each step only
the last two orthogonal polynomials have to be stored in memory. That
is, the memory required for the calculations depends only on the
number of discretization points $N$ but is independent of the order of
approximation. Secondly, no multiprecision arithmetics is needed. All
the calculations can be carried out using the built in 10 Byte
floating point type. Finally, since the integrals are evaluated
numerically, the indefinite integrals of the integrands are not
needed. Therefore, there are no restrictions on the form of the
function $f$ and the weight function $w$.

\subsection{Stability, optimal weight function and discretization}
\label{sec:stab}

In order to describe the numerical stability of the recursion formula
(\ref{eqn:rek}) and to find the optimal weight function $w$ and the
optimal type of discretization we need to refer to $\Lab$, the Hilbert
space of $\left[a,b\right] \to \mathbb{C}\,$ square-integrable
functions with the inner product
$\scalar{f}{g}=\int_a^bf(x)^*g(x)\,\mathrm{d}x$.  If the weight
function $w$ is such that
\begin{equation}
\label{eqn:w}
0<\kappa_1<w(x)<\kappa_2\quad \forall
x\in\left[a,b\right],
\end{equation}
then the equivalence classes in $\Lab$ consist of the same functions
as in $\Lwab$ and $\Lab$ consist of the same equivalence classes as
$\Lwab$. That is, $\Lwab$ and $\Lab$ are identical as linear
spaces.

Both 
\begin{equation}
\Big\{e_n \Bigm| n\in \mathbb{Z}\Big\}
\end{equation}
and
\begin{equation}
\Big\{s_n \Bigm| n\in \mathbb{N}\Big\} \cup \Big\{c_n \Bigm| n\in
\mathbb{N}\Big\} \cup \Big\{c_0\Big\}
\end{equation}
form orthonormal bases in $\Lab$, where
\begin{equation}
\label{eqn:en}
e_n(x)=\frac{1}{\sqrt{b-a}} \exp\left[ \frac{2\pi n i}{b-a} \left( x-
\frac{a+b}{2} \right)\right], \quad\quad n=\dots,-2,-1,0,1,2,\dots
\end{equation}
and
\begin{eqnarray}
\label{eqn:sncn}
s_n(x)&=&\sqrt{\frac{2}{b-a}} \sin\left[ \frac{2\pi n}{b-a} \left( x-
\frac{a+b}{2} \right)\right], \quad\quad n=1,2,3,\dots\nonumber \\
c_n(x)&=&\sqrt{\frac{2}{b-a}} \cos\left[ \frac{2\pi n}{b-a} \left( x-
\frac{a+b}{2} \right)\right], \quad\quad n=1,2,3,\dots \\
c_0(x)&=&\frac{1}{\sqrt{b-a}}=e_0(x). \nonumber
\end{eqnarray}

\begin{figure}
\begin{center}
 \resizebox{80mm}{!}{\includegraphics{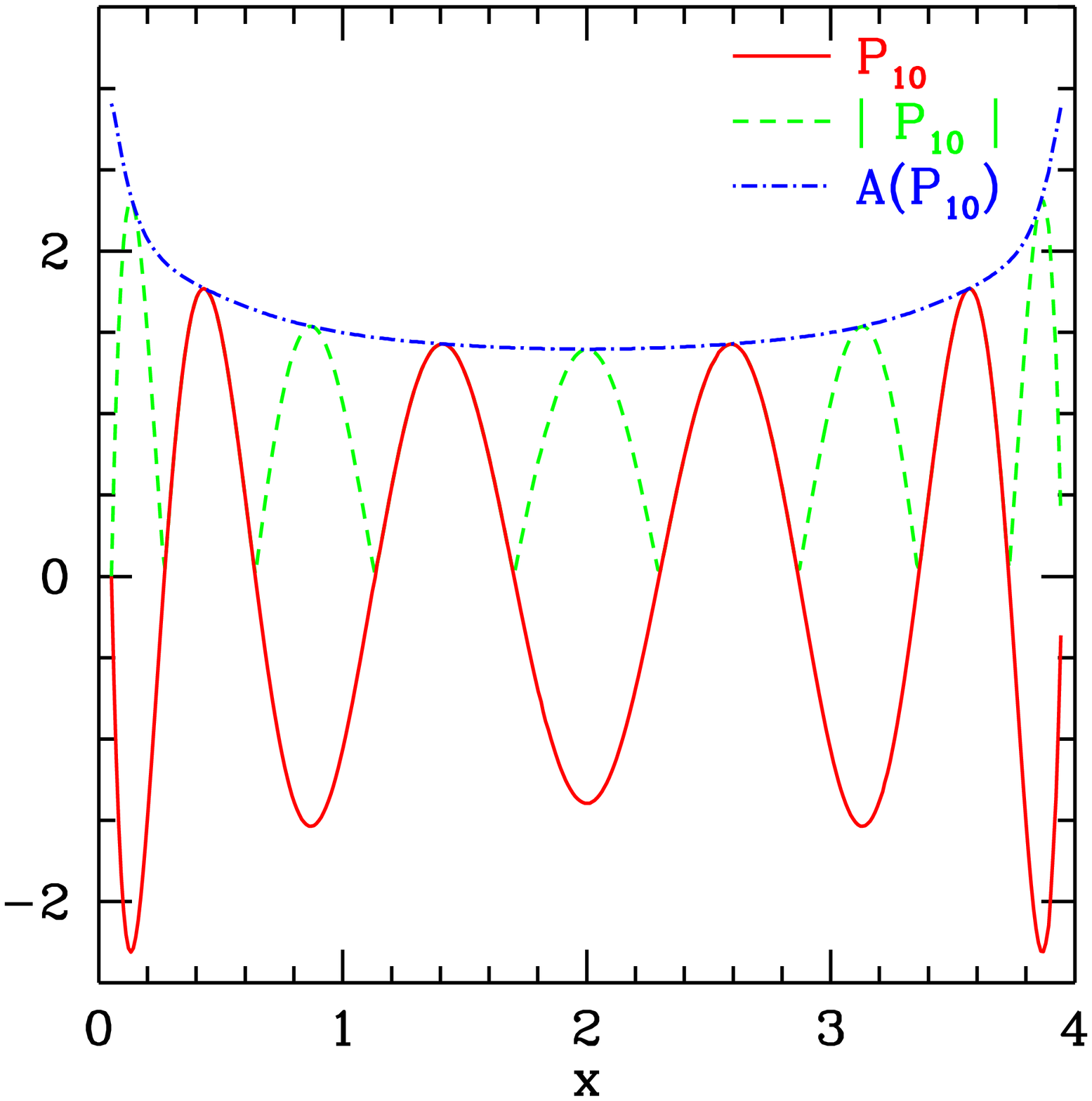}}
 \resizebox{80mm}{!}{\includegraphics{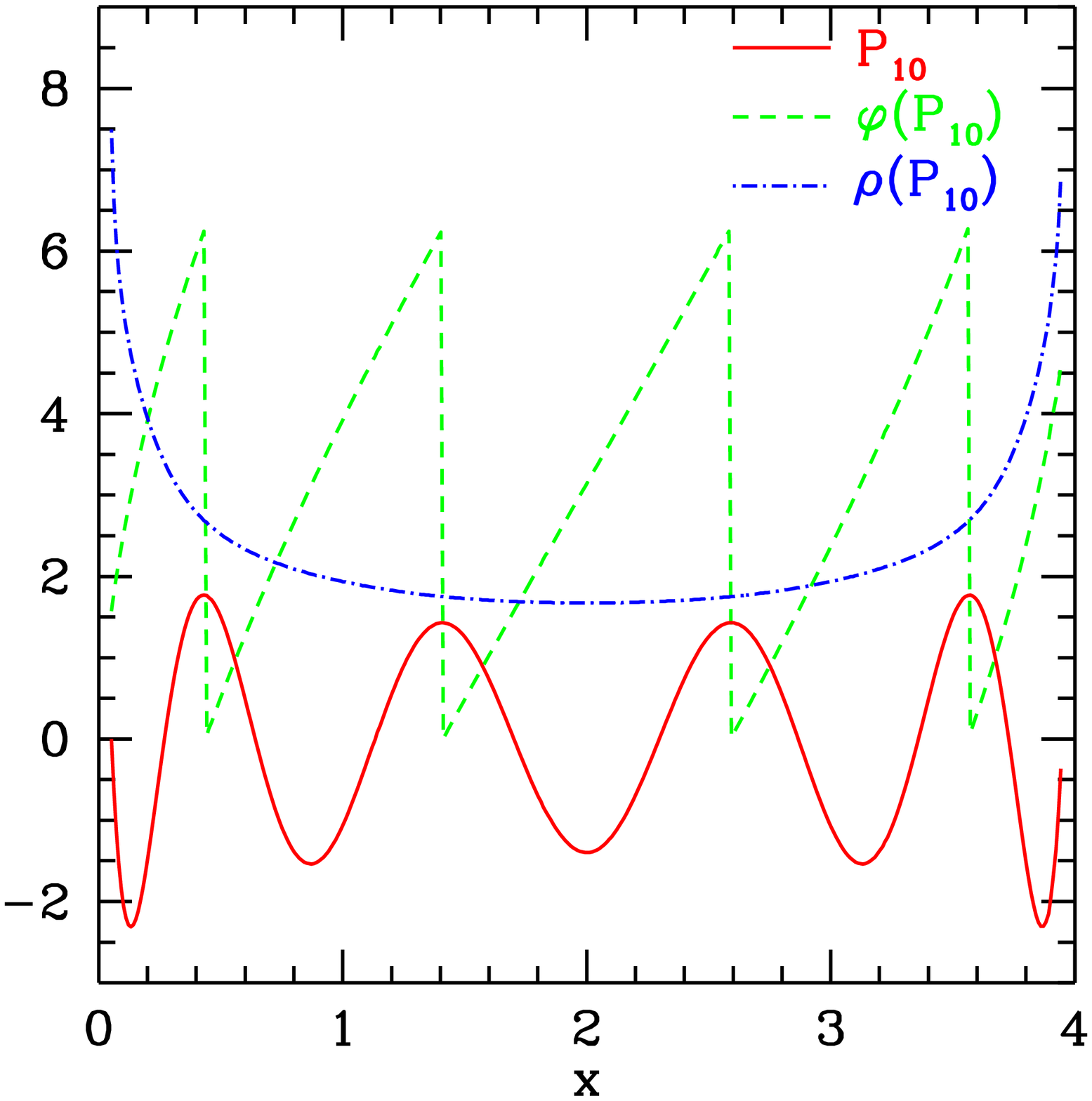}}
\caption{The amplitude (left), phase and density (right) of the 10th
orthogonal polynomial $P_{10}$ with respect to the weight function $w=1$
on the interval $\left[0,4\right]$.}
\label{fig:amp}
\end{center}
\end{figure}

Using these basis vectors the linear map 
\begin{equation}
\label{eqn:F}
F := I + i\cdot\ssum_{n=1}^{\infty} \Big( \ketbra{s_n}{c_n} -
\ketbra{c_n}{s_n} \Big) = \ketbra{e_0}{e_0} + 2 \cdot\ssum_{n=1}^{\infty}
\ketbra{e_n}{e_n}
\end{equation}
can be defined, which is bounded and self-adjoint in $\Lab$. Let
\begin{equation}
\label{eqn:A}
A_f(x):=\big| \big(Ff\big)(x) \big|
\end{equation}
and
\begin{equation}
\label{eqn:phi}
\varphi_f(x):=\arg \bigg(\big(Ff\big)(x)\bigg).
\end{equation}
Then
\begin{equation}
\big(Ff\big)(x)=A_f(x)\cdot \exp\left(i\varphi_f(x)\right).
\end{equation}
If $f$ is real, then
\begin{equation}
\label{eqn:freal}
f(x)=\Ree \bigg(\big(Ff\big)(x)\bigg)=A_f(x)\cdot
\cos\left(\varphi_f(x)\right).
\end{equation}
That is, $A_f(x)$ and $\varphi_f(x)$ can be naturally identified as the
\emph{amplitude} and \emph{phase} of the real function $f$ at point
$x$, respectively.
If $\varphi_f$ is differentiable then we can define
\begin{equation}
\label{eqn:rho}
\varrho_f(x):=\frac{1}{\pi}\varphi_f'(x).
\end{equation}
If $f$ is a polynomial without multiple zeros (which is the case when
dealing with orthogonal polynomials) then $\varphi_f$ is strictly
increasing and $\varrho_f>0$. In this case if
\begin{equation}
\Int{\varrho_f(x)}{x}{x_1}{x_2}=k
\end{equation}
for some $a<x_1<x_2<b$, then $f$ has exactly $k$ zeros in both
$\left[x_1,x_2\right[$ and $\left]x_1,x_2\right]$. Thus, $\varrho_f$
can be identified as the \emph{density} or \emph{root density} of
polynomial $f$. Graphically speaking, $A_f(x)$ describes the
'amplitude' of the polynomial $f$ at point $x$ and $\varrho_f(x)$
describes the 'rate at which the polynomial $f$ oscillates' near $x$
(Figure \ref{fig:amp}).  If $f_1$ and $f_2$ are polynomials without
multiple zeros such that $f_1$ and $f_2$ have no common zeros, then
the number of roots of $f_1\cdot f_2$ in every $\left[ x_1,x_2 \right]
\subset \left[a,b\right]$ is equal to the sum of the number of roots
of $f_1$ and $f_2$ in that interval. Therefore, the root density of
such a product is approximately the sum of the root density of the
factors.

Let $\Phi_\mu$ and $P_\mu$ denote the $\mu$th orthogonal polynomials
in $\Lwab$ and $\Lab$, respectively. If $a=-1$ and $b=1$ then the
polynomials $P_\mu$ are equal to the Legendre polynomials. Using the
formulae for the asymptotic behaviour of the Legendre polynomials
\cite[\S{}8.21]{szego} the formula
\begin{equation}
\label{eqn:amp1}
A_{P_{\mu}}(x) = \text{const}\cdot \frac{1}{\D \sqrt[4]{ \left(
\frac{b-a}{2} \right)^2 - \left(x- \frac{a+b}{2} \right)^2}} + O(\mu^{-3/2})
\end{equation}
can be obtained for the amplitude $A_{P_\mu}$ of the polynomials
$P_\mu$ for large $\mu$s.

\begin{figure}
\begin{center}
 \resizebox{80mm}{!}{\includegraphics{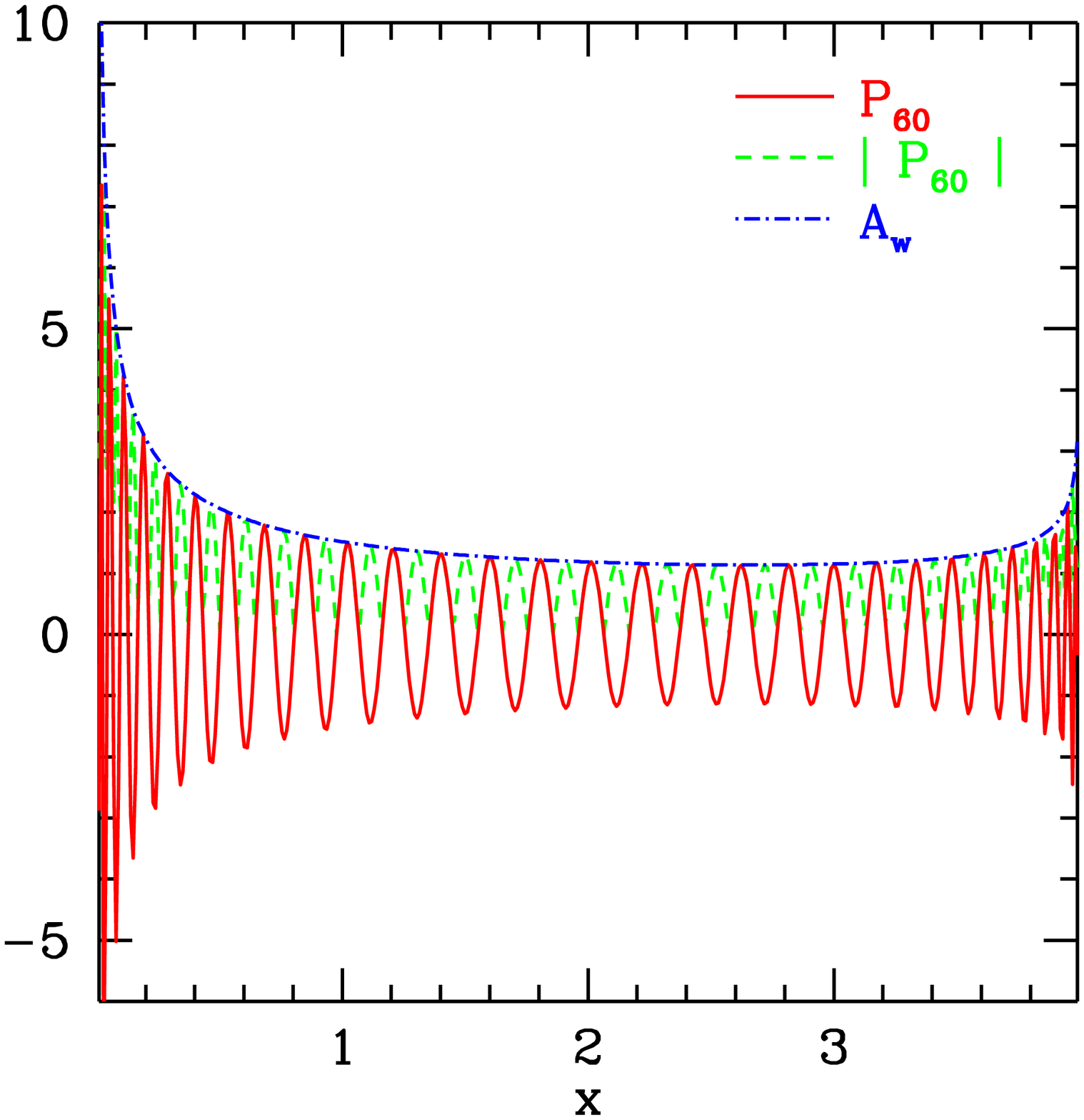}}
 \resizebox{80mm}{!}{\includegraphics{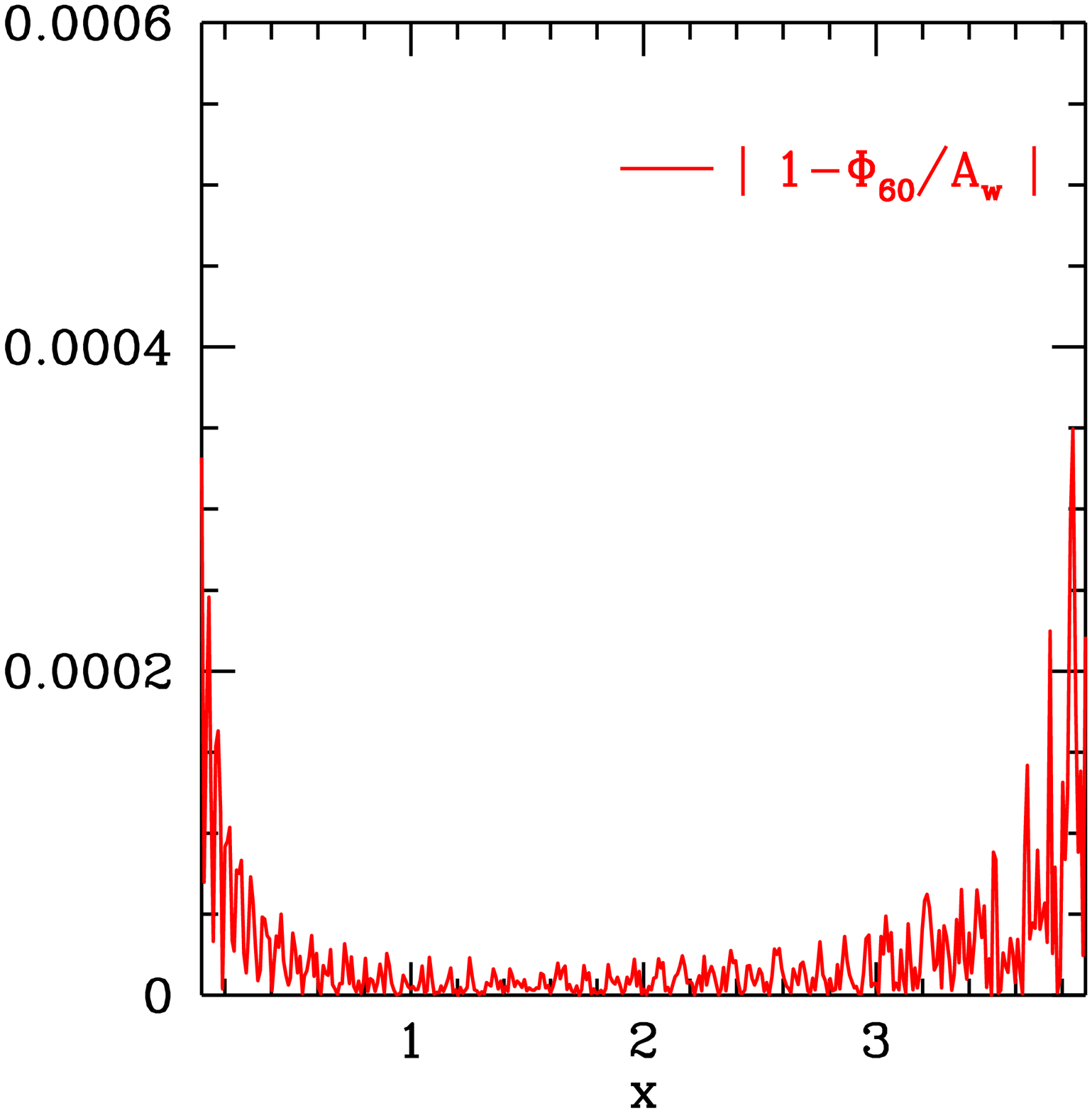}}
\caption{Left: $A_w$, defined in equation (\ref{eqn:amp}), as the
  amplitude of $\Phi_{60}$. Right: The relative deviation of the
  amplitude of $\Phi_{60}$ and $A_w$ with $a=10^{-5}$, $b=4$ and
  $w(x)=x^{-1/4}$.  (The amplitude of $\Phi_{60}$ was calculated
  numerically using only the first 5000 terms of the sum in equation
  (\ref{eqn:F}).)}
\label{fig:amp2}
\end{center}
\end{figure}

If $w$ satisfies condition (\ref{eqn:w}), then
\begin{equation}
w(x) A_{\Phi_\mu}(x)\approx \text{const}\cdot A_{P_\mu}(x)
\end{equation}
for large $\mu$s, that is, for large $\mu$s the amplitude of
$\Phi_\mu$ can be well approximated by
\begin{equation}
\label{eqn:amp2}
A_{\Phi_\mu}(x) \approx \text{const} \cdot \frac{1}{w(x)} A_{P_\mu}(x),
\end{equation}
where the constant is independent of $x$ and is near $1$ (Figure
\ref{fig:amp2}). Combining equations (\ref{eqn:amp1}) and
(\ref{eqn:amp2}) the amplitude of the $\mu$th orthogonal polynomial
$\Phi_\mu$ can be approximately given by the formula
\begin{equation}
\label{eqn:amp}
A_{\Phi_\mu}(x) \approx \text{const}\cdot \frac{1}{w(x)}\,
\frac{1}{\D \sqrt[4]{ \left(
\frac{b-a}{2} \right)^2 - \left(x- \frac{a+b}{2} \right)^2}} =: A_w(x).
\end{equation}

\begin{figure}
\begin{center}
 \resizebox{90mm}{!}{\includegraphics{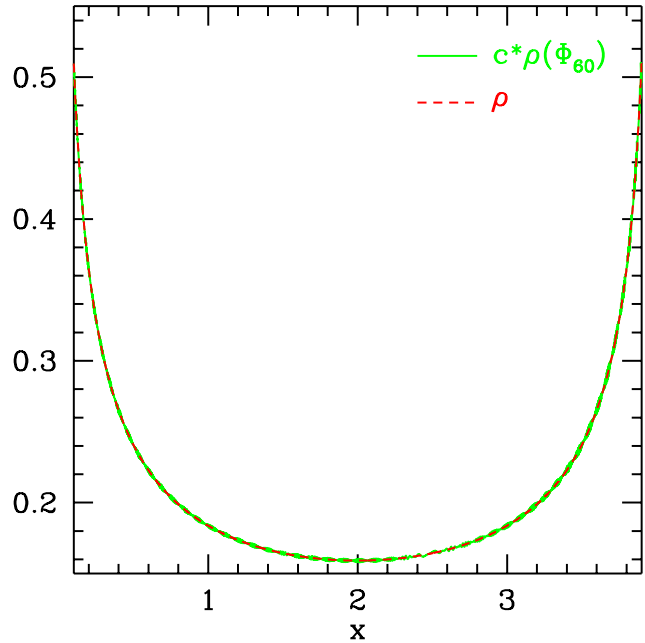}}
\caption{$\frac{1}{60}\cdot 0.9875 \cdot\varrho_{\Phi_{60}}(x)$
  calculated numerically (solid line) such that only the first 5000
  terms of the sum in equation (\ref{eqn:F}) were taken into account
  and $\varrho(x)$ (dashed line) with $a=10^{-5}$, $b=4$ and
  $w(x)=x^{-1/4}$.}
\label{fig:rho}
\end{center}
\end{figure}

For large $\mu$s the root density of the polynomials can be
approximated by
\begin{equation}
\label{eqn:ro}
\varrho_{\Phi_\mu}(x) \approx
\mu\,\omega_{\mu}\cdot\frac{\D\frac{1}{\pi}}{\D\sqrt{ \left(
\frac{b-a}{2} \right)^2 -
\left(x- \frac{a+b}{2} \right)^2}} =: \mu\,\omega_\mu\cdot\varrho(x),
\end{equation}
where $\omega_\mu$ depends on the weight function $w$. In all
cases $\omega_\mu$ is close to 1 and $\D \lim_{\mu\to\infty} \omega_\mu
= 1$ (Figure \ref{fig:rho}).

\begin{figure}
\begin{center}
 \resizebox{80mm}{!}{\includegraphics{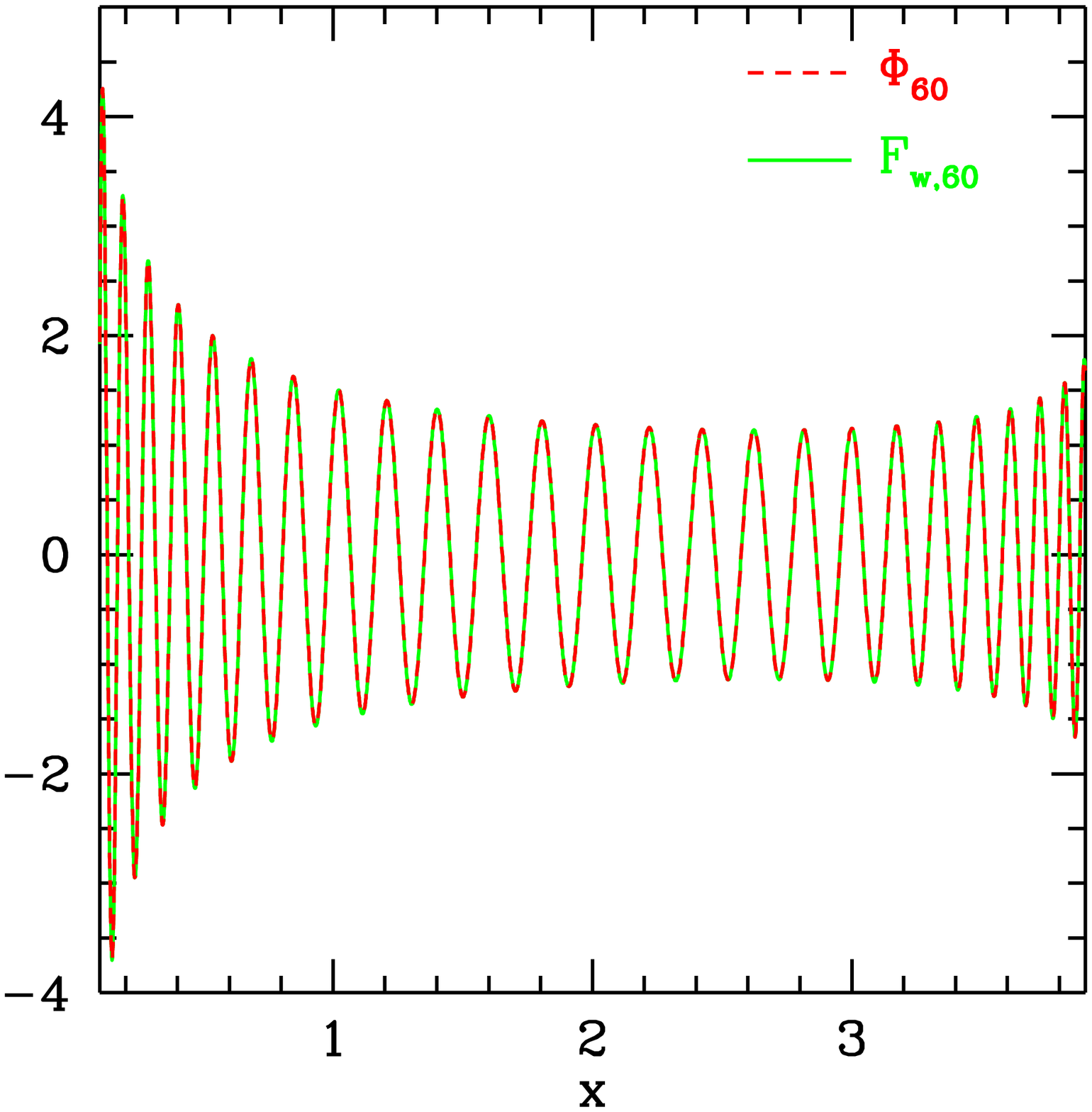}}
 \resizebox{80mm}{!}{\includegraphics{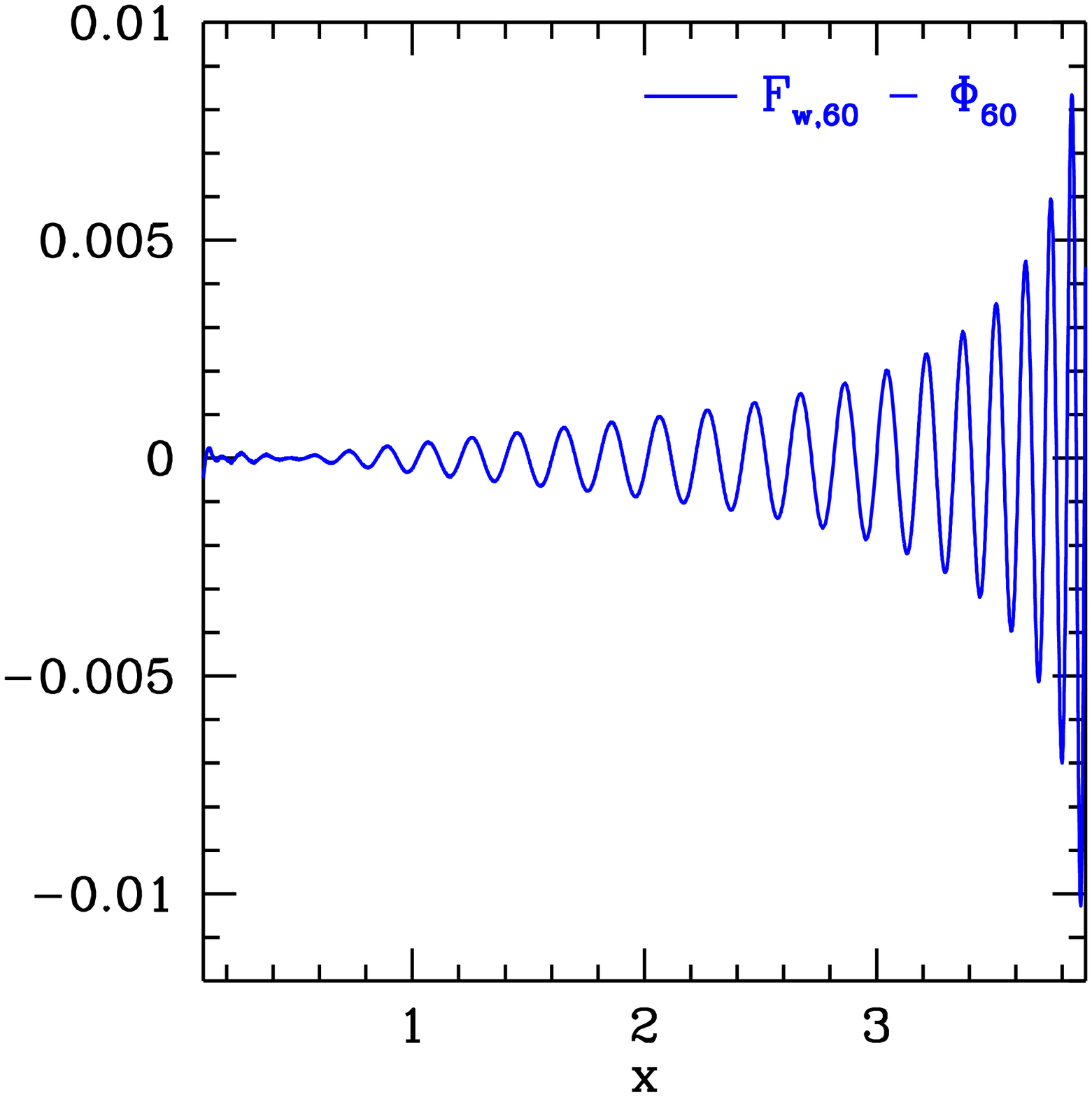}}
\caption{Left: $F_{w,60}$ (solid line) with $\omega_{60}=0.9875$ and
  $\Phi_{60}$ (dashed line). Right: The difference of $F_{w,60}$ and
  $\Phi_{60}$. ($a=10^{-5}$, $b=4$ and $w(x)=x^{-1/4}$)}
\label{fig:approx}
\end{center}
\end{figure}

With
\begin{equation}
\label{eqn:varphi}
\varphi(x):= \pi\Int{\varrho(z)}{z}{a}{x}=\pi-\arccos
\left(\frac{2x-a-b}{b-a} \right)
\end{equation}
and using equations (\ref{eqn:freal}), (\ref{eqn:amp}) and
(\ref{eqn:ro}) we can conclude at
\begin{equation}
\Phi_{\mu}(x) \approx A_w(x) \cdot \cos\left\{
\mu\,\omega_\mu\left[\pi - \arccos \left(\frac{2x-a-b}{b-a} \right)
\right] \right\}=:F_{w,\mu}(x),
\end{equation}
an approximate formula for the $\mu$th orthogonal polynomial with
respect to the weight fuction $w$ (Figure \ref{fig:approx}).

We use Simpson's rule in order to calculate the integrals
(\ref{eqn:q1}), (\ref{eqn:b}) and (\ref{eqn:p}) numerically. The
numerical integral of the function $g$ taken over the discretization
interval $I=\left[z-h,z+h\right]$ is
\begin{equation}
\Sim{I} g = \left[\frac13 g(z-h) + \frac43 g(z) +
\frac13 g(z+h)\right]h.
\end{equation}
Assume that $g$ is analytic in $I$ with radius of convergence at $z$
greater then $h$, that is,
\begin{equation}
g(z+u)=\sum_{n=0}^{\infty} \frac{g^{(n)}(z)}{n!}\,u^n\quad\quad -h\le
u \le h.
\end{equation}
The exact integral and the numerical integral of $g$ then becomes
\begin{equation}
\int_I g = \sum_{n=0}^{\infty}
\frac{g^{(n)}(z)}{n!}\,\frac{h^{n+1}}{n+1} - 
\sum_{n=0}^{\infty} \frac{g^{(n)}(z)}{n!}\,\frac{(-h)^{n+1}}{n+1}=
2\sum_{k=0}^{\infty}\frac{g^{(2k)}(z)}{(2k+1)!}\,h^{2k+1}
\end{equation}
and
\begin{eqnarray}
\Sim{I} g&=&\left[\frac13
\sum_{n=0}^{\infty}\frac{g^{(n)}(z)}{n!}\,(-h)^n + 
\frac43 \sum_{n=0}^{\infty}\frac{g^{(n)}(z)}{n!}\,(0)^n +
\frac13 \sum_{n=0}^{\infty}\frac{g^{(n)}(z)}{n!}\,h^n 
\right]h= \nonumber\\
&=&\left[
\frac23 \sum_{k=0}^{\infty}\frac{g^{(2k)}(z)}{(2k)!}\,h^{2k} + \frac43
g(z) \right]h,
\end{eqnarray}
respectively. Then the error of integration over the interval $I$, that
is, the difference of the exact and the numerical integrals becomes
\begin{equation}
\Delta_I g = \left| \int_I g - \Sim{I} g\right| = \left|
\sum_{k=2}^{\infty}
\frac{g^{(2k)}(z)}{(2k)!}\,\left(\frac23-\frac{2}{2k+1}\right)
h^{2k+1} \right|.
\label{eqn:Deltaf1}
\end{equation}
If $h$ is small enough and $\mu$ is large enough such that $\Phi_\mu$
oscillates much faster than $w$ and $f$ the integrand of (\ref{eqn:b})
can be approximated on $I$ by
\begin{equation}
F_z(u):=B(z)\cos\left(\pi\,\mu\,\varrho(z)\cdot u+ \varphi_{\Phi_\mu}(z)
\right)\quad\quad -h\le u \le h,
\end{equation}
where $B(z)=w(z)^2f(z)A_{\Phi_\mu}(z)$.
\begin{equation}
F_z^{(2k)}(u)=(-1)^k\left(\pi\,\mu\,\varrho(z)\right)^{2k}\cdot
F_z(u),
\end{equation}
thus, the error of integrating $F_z$ numerically over $I$ can be
estimated as
\begin{eqnarray}
\Delta_I F_z &=& \left| F_z(z)\right|\cdot \left| \sum_{k=2}^{\infty}
\frac{(-1)^k \left( \pi\,\mu\,\varrho(z)\right)^{2k}}
{(2k)!}\,\left(\frac23-\frac{2}{2k+1}\right) h^{2k+1} \right| \le
\nonumber  \frac23 h \left| F_z(z) \right| \sum_{k=2}^{\infty} \frac{
\left( \pi\,\mu\,\varrho(z)\right)^{2k}h^{2k}}{(2k)!} \\
&\le&\frac23 h \left|B(z)\right| \cdot
\left[\cosh\left(\pi\,\mu\,\varrho(z)
\,h\right) - 1 - \frac{\left(\pi\,\mu\,\varrho(z)
\,h\right)^2}{2}\right] \nonumber \\
&\approx&
\frac{1}{36}h\left|B(z)\right| \left(\pi\,\mu\,\varrho(z)
\,h\right)^4 \cdot \left[ 1 + O\left(\left(\pi\,\mu\,\varrho(z)
\,h\right)^2\right) \right].
\label{eqn:Deltaf2}
\end{eqnarray}
$1/\left(\pi\,\mu\,\varrho(z)\,h\right)$ shows the number of
discretization points between two consecutive zeros of $\Phi_\mu$ near
$z$. The density of discretization has to be chosen such that
the relative error
\begin{equation}
\delta_I F_z = \Frac{\Delta_I F_z}{\left| B(z) \right|\cdot h}
\end{equation}
of the numerical integration is in the same order of magnitude in each
discretization interval and is equal to the desired relative error of
the numerical integration over $\left[a,b\right]$.

Since $\cos^2 x = (1+\cos 2x)/2$ and the integrands in
(\ref{eqn:q1}) and (\ref{eqn:p}) contain the square of $\Phi_\mu$,
they can be treated as a constant plus a cosine of double frequency in
every discretization interval. The error of the numerical integration
can be estimated similarly, leading us to the same conclusion.

If $f$ is not continuous on $\left[a,b\right]$, then the convergence
in (\ref{eqn:f=}) is not uniform but an almost everywhere convergence.
Assume that $f$ has an $(x-a)^{-\alpha}$ type ($\alpha>0$) singularity
near $a$, that is, $\D 0<\left| \lim_{x\to a^+} f(x)(x-a)^\alpha
\right| < \infty$. Then the closer we get to $a$ the slower the
convergence becomes. Thus, by taking into account only a finite number
of terms in (\ref{eqn:f=}) we can get a reasonable approximation for
$f$ only in $\left[ a+\epsilon,b\right]$ with a suitably chosen
$\epsilon$. The procedure in this case should be as follows. We need
to determine the size $\epsilon$ of the neighborhood of $a$ in which
the approximation of $f$ is not needed. Then we generate the
orthogonal polynomials using (\ref{eqn:rek}) in the interval
$\left[a'=a+\epsilon,b\right]$ and calculate the approximating
polynomial $P_f^{(n)}$ with the desired degree of approximation $n$.

If $f$ does not have singularities and is continuous on
$\left[a,b\right]$ then the weight function $w$ can be chosen to be an
arbitrary continuous function. If $f$ has singularities, for example
an $(x-a)^{-\alpha}$ type singularity near $a$, then the best choice
for $w$ is as follows. $w$ should have an $(x-a)^\alpha$ type
behaviour near $a$ and should be a smooth function otherwise. If $f$
is such that $\left| f(x) \right| > \kappa > 0\,\,\, \forall
x\in\left] a,b \right]$ then the best choice is $w(x)=1/\left| f(x)
\right|$. Choosing $w$ in such a way has the following advantages.
1.~According to (\ref{eqn:amp2}) the amplitude of the polynomials
$\Phi_\mu$ will gain approximately the same type of singularity as $f$
has, therefore, the relative deviation of $f$ and $P_f^{(n)}$ is
decreasing uniformly. 2.~In the integrand of equation (\ref{eqn:b})
the singularity of $f$ is cancelled out by one of the two $w$'s, while
the other $w$ deals with $\Phi_\mu$ according to (\ref{eqn:amp2}).

In order to find the optimal type of discretization of interval
$\left[a,b \right]$ we need to consider both the singularities and the
densities of the integrands (\ref{eqn:q1}), (\ref{eqn:p}) and
(\ref{eqn:b}). Taking (\ref{eqn:amp1}) and (\ref{eqn:amp2}) into
account we can conclude that $\left| B(x) \right|$ of (\ref{eqn:q1})
and (\ref{eqn:p}) (see (\ref{eqn:Deltaf2})) has singularities near $a$
and $b$ of type $1/\sqrt{x-a}$ and $1/\sqrt{b-x}$, respectively, and
(\ref{eqn:b}) has singularities of type $1/\sqrt[4]{x-a}$ and
$1/\sqrt[4]{b-x}$. The densities of all three integrands are of type
$1/\sqrt{(x-a)(b-x)}$. As a consequence, the optimal discretization
should contain the discretization points with density proportional to
$1/[(x-a)(b-x)]$. This would require infinitely many discretization
points, thus, we choose a small $\epsilon$ and discretize the interval
$I'= \left[a',b'\right]= \left[a+\epsilon, b-\epsilon \right]$ with
the above density. If we use $N$ discretization points, then the
discretization interval $\left[z-h_z, z+h_z \right]$ at $z$ has the
length
\begin{equation}
2h_z=\left\{
\begin{matrix}
\D (z-a)\left[ \left( \frac{b-a}{2\epsilon} \right)^{4/N} -1 \right]
\quad\quad\text{if} \quad a+\epsilon \le z \le \frac{a+b}{2} \\ 
\D (b-z)\left[ \left( \frac{b-a}{2\epsilon} \right)^{4/N}-1 \right]
 \quad\quad \text{if} \quad\frac{a+b}{2} \le z \le b-\epsilon. \\
\end{matrix} \right.
\end{equation}
The length of the longest and shortest intervals of this logarithmic
discretization is $\frac{\epsilon}{2}\left[ \left(
\frac{b-a}{2\epsilon} \right)^{4/N} -1 \right]$ and $\frac{b-a}{4}
\left[ \left( \frac{b-a}{2\epsilon} \right)^{4/N}-1 \right]$,
respectively.

In the usual fermionic calculations $a=0$ and the functions of the
fermionic matrix that have to be evaluated have singularities of type
$x^{-\alpha}$ ($\alpha > 0$) at $x=0$. We use the above polynomial
expansion to approximate these singular functions. $\epsilon$ should
be chosen such that all eigenvalues of the fermionic matrix are
greater then $\epsilon$. The smallest eigenvalue of the fermionic
matrix is proportional to the square of the quark mass.  Since
our aim is to use quark masses as low as the physical $u,d$ quark
masses, which are approximately $m_{ud}=0.002$ in lattice mass units,
$\epsilon$ should be in the order of magnitude of $10^{-6}$. In order
to be able to well approximate the functions of the fermionic matrix
so close to their singularities the required order of polynomial
approximation is in the thousands.

\begin{figure}
\begin{center}
 \resizebox{80mm}{!}{\includegraphics{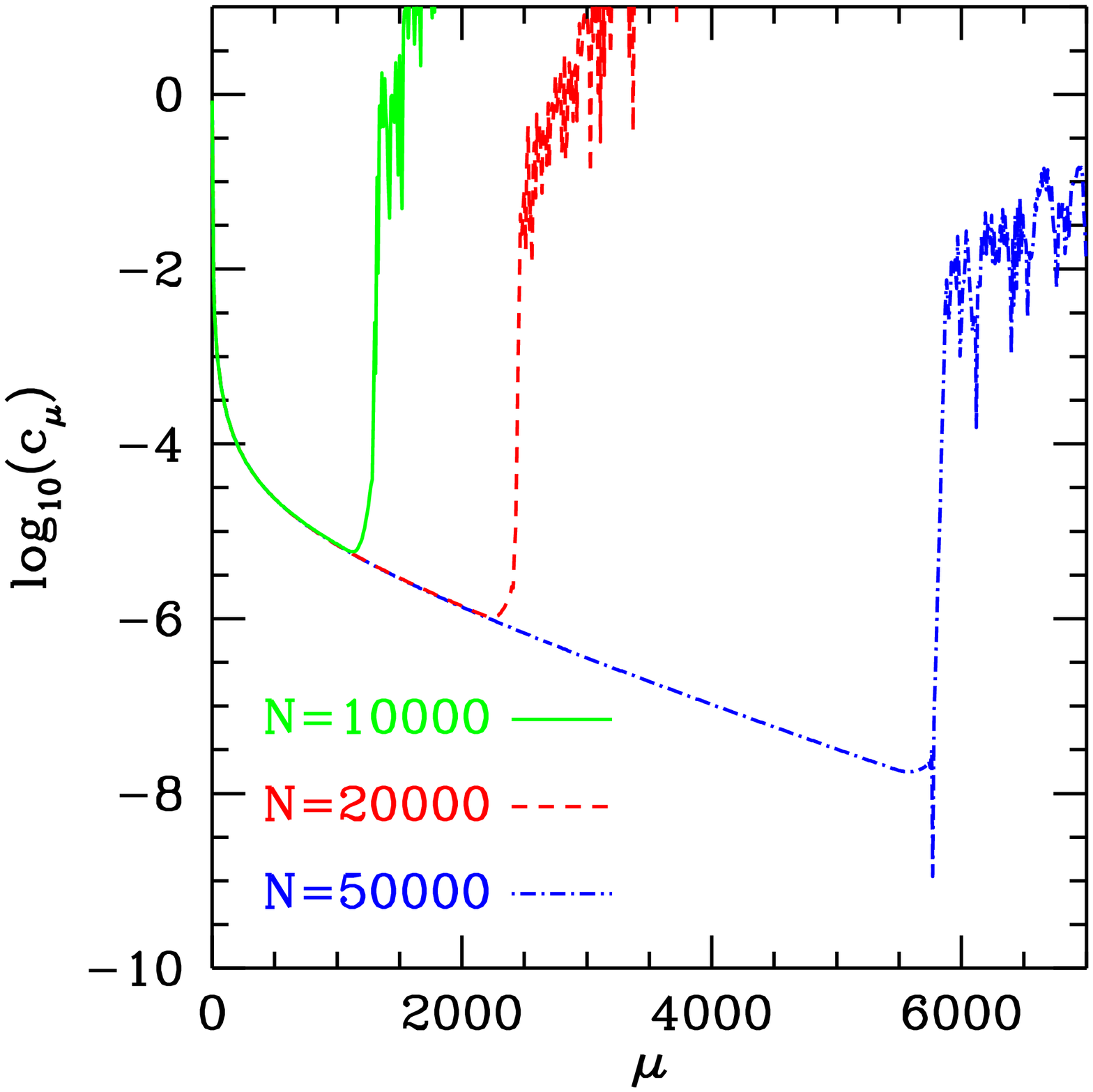}}
 \resizebox{80mm}{!}{\includegraphics{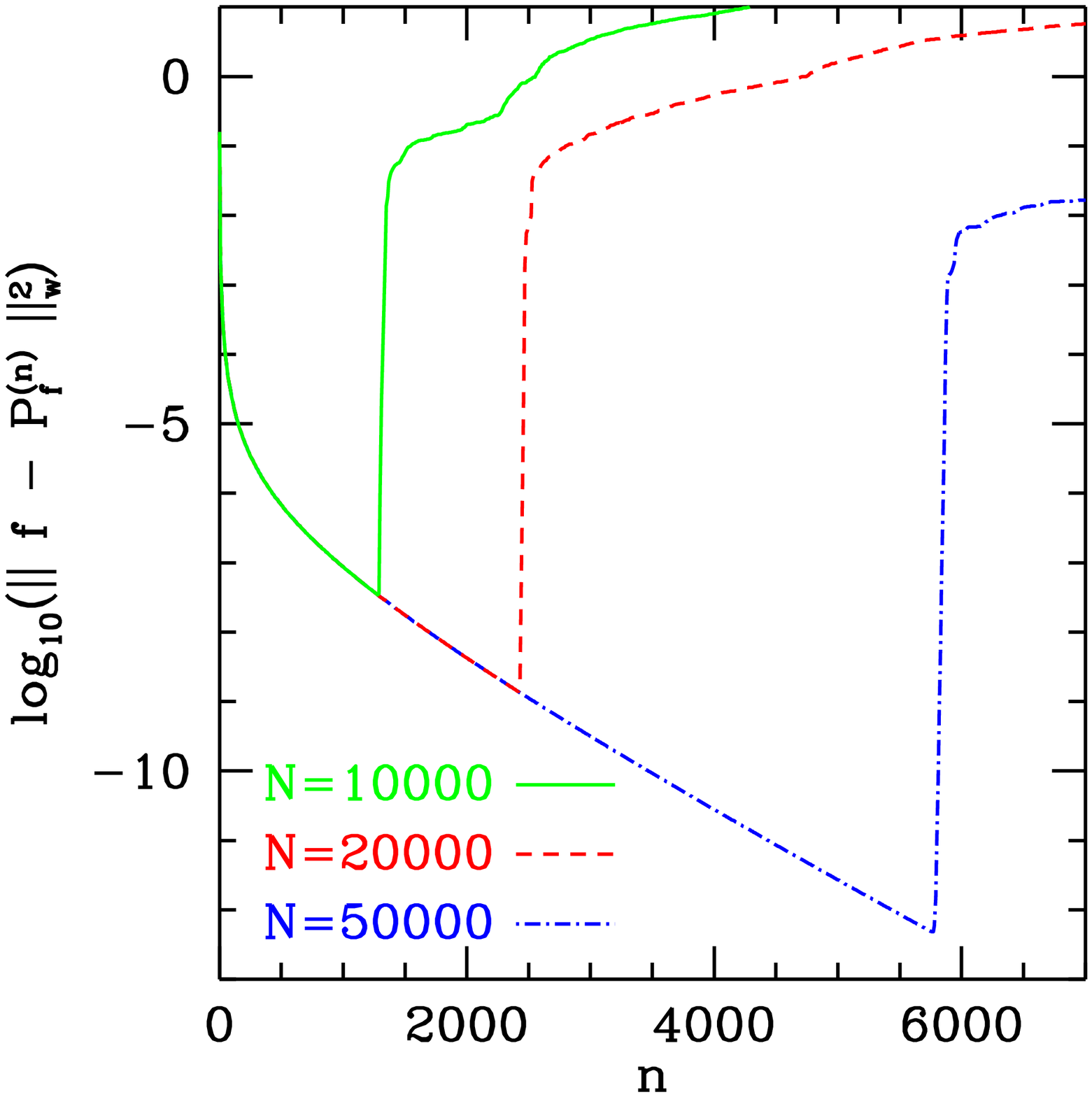}}
\caption{The logarithm of the coefficients $c_\mu$ (left) and the
  deviations $\normw{f-P_f^{(n)}}^2$ (right) as the function of the
  order with number of discretization points $N=10000$, $N=20000$ and
  $N=50000$.}
\label{fig:coeff}
\end{center}
\end{figure}

Using this logarithmic type of discretization the order $n$ up to
which the algorithm (\ref{eqn:rek}) is stable can be tested. The
coefficients $c_\mu$ and the deviations $\normw{f-P_f^{(n)}}^2$ can be
seen in Figure \ref{fig:coeff}, when the approximated function is
chosen to be $f=x^{-1/4}$, $a=0$, $b=4$ and $\epsilon=10^{-6}$. It can
be verified that the algorithm is stable approximately up to the
orders $n=1000$, $n=2000$ and $n=5500$ if the numbers of
discretization points are $N=10000$, $N=20000$ and $N=50000$,
respectively. Since no multiprecision arithmetics is required for the
algorithm, its CPU time and memory requirements are considerably low.
Generating the polynomials up to the order of $n=5500$ in the case of
$N=50000$ takes approx.~2.5 minutes of CPU time on a $1.6\,\text{GHz}$
AMD Athlon processor and requires approx.~$5\,\text{MB}$ of RAM.

\section{Conclusion}
\label{sec:conc}

We have introduced an alternate numerical method for generating the
approximating polynomials used in fermionic calculations with smeared
link actions. This algorithm was based on the idea of calculating all
the integrals numerically and calculating and storing all the
functions and polynomials only at finitely many discretization points.
The advantages of this algorithm include memory usage independent of
the order of approximation, unnecessarity of multiprecision
arithmetics libraries and the absence of restrictions for the form of
the approximated and the weight functions. We investigated the
stability of the algorithm and based on the asymptotic behaviour of
the orthogonal polynomial base appearing in the method we determined
the optimal weight function and the optimal type of discretization. As
a result the achievable order of polynomial approximation reached
several thousands which is essential for fermionic calculations near
the small physical quark masses.

\section{Acknowledgments}
We thank Ferenc Csikor, Zolt\'an Fodor and Istv\'an Montvay for useful
discussions and careful reading of the manuscript. This work was partially
supported by Hungarian Scientific grants, OTKA-T37615/T34980.


\begin{thebibliography}{99}

\bibitem{9809157}
G.~P.~Lepage,
Phys.\ Rev.\ D {\bf 59}, 074502 (1999) [arXiv:hep-lat/9809157].

\bibitem{9903032}
K.~Orginos, D.~Toussaint and R.~L.~Sugar  [MILC Collaboration],
Phys.\ Rev.\ D {\bf 60}, 054503 (1999)
[arXiv:hep-lat/9903032].

\bibitem{0103029}
A.~Hasenfratz and F.~Knechtli,
Phys.\ Rev.\ D {\bf 64}, 034504 (2001)
[arXiv:hep-lat/0103029].


\bibitem{Kamleh:2003wb}
W.~Kamleh, D.~B.~Leinweber and A.~G.~Williams,
arXiv:hep-lat/0309154.

\bibitem{Zanotti:2001yb}
J.~M.~Zanotti {\it et al.}  [CSSM Lattice Collaboration],
Phys.\ Rev.\ D {\bf 65} (2002) 074507
[arXiv:hep-lat/0110216].

\bibitem{Morningstar:2003gk}
C.~Morningstar and M.~J.~Peardon,
arXiv:hep-lat/0311018.




\bibitem{0012022}
F.~Knechtli and A.~Hasenfratz,
Phys.\ Rev.\ D {\bf 63}, 114502 (2001)
[arXiv:hep-lat/0012022].

\bibitem{0203010}
A.~Hasenfratz and F.~Knechtli,
Comput.\ Phys.\ Commun.\  {\bf 148}, 81 (2002)
[arXiv:hep-lat/0203010].

\bibitem{0203026}
A.~Hasenfratz and A.~Alexandru,
Phys.\ Rev.\ D {\bf 65}, 114506 (2002)
[arXiv:hep-lat/0203026].

\bibitem{Hasenbusch:1998yb}
M.~Hasenbusch,
Phys.\ Rev.\ D {\bf 59}, 054505 (1999)
[arXiv:hep-lat/9807031].

\bibitem{deForcrand:1998sv}
P.~de Forcrand,
Nucl.\ Phys.\ Proc.\ Suppl.\  {\bf 73} (1999) 822
[arXiv:hep-lat/9809145].



\bibitem{9903029}
I.~Montvay,
arXiv:hep-lat/9903029.

\bibitem{9911014}
I.~Montvay,
arXiv:hep-lat/9911014.

\bibitem{0302025}
C.~Gebert and I.~Montvay,
arXiv:hep-lat/0302025.

\bibitem{szego} 
Gabor Szeg\H{o}: \emph{Orthogonal Polynomials}, American Mathematical
Society Colloquium Publications, Vol 23, 4th Ed, Providence, RI, 1975

\end{thebibliography}
\end{document}